\documentclass[twocolumn,prl,showpacs]{revtex4}
\usepackage{stmaryrd}
\usepackage{amsmath}
\usepackage{amssymb}
\usepackage{graphicx}
\usepackage{dcolumn}
\usepackage{bm}

\begin{document}

\begin{titlepage}

%\title{Orbital-selective symmetry breaking leads to Dirac half-metal and quantum anomalous Hall system: the case of Mn intercalated epitaxial graphene}
\title{Theory of the Dirac Half Metal and Quantum Anomalous Hall Effect in Mn Intercalated Epitaxial Graphene}

\author{Yuanchang Li$^1$, Damien West$^2$\footnote{damienwest@gmail.com}, Huaqing Huang$^3$, Jia Li$^4$, S. B. Zhang$^2$, and Wenhui Duan$^{3,5}$\footnote{dwh@phys.tsinghua.edu.cn}}
\address{$^1$National Center for Nanoscience and Technology, Beijing 100190, People��s Republic of China }
\address{$^2$Department of Physics, Applied Physics, and Astronomy, Rensselaer Polytechnic Institute, Troy, New York 12180, USA}
\address{$^3$Department of Physics and State Key Laboratory of Low-Dimensional Quantum Physics, Tsinghua University, Beijing 100084, People's Republic of China}
\address{$^4$Institute of Advanced Materials, Graduate School at Shenzhen, Tsinghua University, Shenzhen 518055, People's Republic of China}
\address{$^5$Collaborative Innovation Center of Quantum Matter, Tsinghua University, Beijing 100084, China}
\date{\today}

\begin{abstract}

%{The prospect of a Dirac half metal, a material which is characterized by a bandstructure with a gap in one spin channel but a Dirac cone in the other, is not only of fundamental interest but a natural candidate for use in spin-polarized current generation applications. However, while the possibility of such a material has been reported based on lattice model studies [H. Ishizuka and Y. Motome, Phys. Rev. Lett. \textbf{109}, 237207 (2012)], it remains unclear what material system might realize such an exotic state.Using first-principles calculations, we show that the experimentally accessible Mn intercalated epitaxial graphene on SiC(0001) transits to a Dirac half metal when the coverage is $>$ 1/3 monolayer. The formation of the Dirac half metal results from an orbital-selective breaking of quasi-2D inversion symmetry, leading to symmetry breaking in a single spin channel. The symmetry breaking is robust, an randomness in the distribution of Mn intercalates preserves the gap opening. Moreover, spin-orbit coupling of Mn drives the system into a quantum anomalous Hall (QAH) state. Our results thus not only demonstrate the practicality of realizing Dirac half metal beyond a toy model but also open a new avenue to the realization of the QAH effect.}
The prospect of a Dirac half metal, a material which is characterized by a bandstructure with a gap in one spin channel but a Dirac cone in the other, is of both fundamental interest and a natural candidate for use in spin-polarized current applications. However, while the possibility of such a material has been reported based on model calculations[H. Ishizuka and Y. Motome, Phys. Rev. Lett. \textbf{109}, 237207 (2012)], it remains unclear what material system might realize such an exotic state. Using first-principles calculations, we show that the experimentally accessible Mn intercalated epitaxial graphene on SiC(0001) transits to a Dirac half metal when the coverage is $>$ 1/3 monolayer. This transition results from an orbital-selective breaking of quasi-2D inversion symmetry, leading to symmetry breaking in a single spin channel which is robust against randomness in the distribution of Mn intercalates. Furthermore, the inclusion of spin-orbit interaction naturally drives the system into the quantum anomalous Hall (QAH) state. Our results thus not only demonstrate the practicality of realizing the Dirac half metal beyond a toy model but also open up a new avenue to the realization of the QAH effect.
\end{abstract}


\maketitle

\end{titlepage}
The discoveries of novel band structures, such as the Dirac spectrum\cite{graphene,ZhouSY, Kane,konig,ZhangHJ} and half-metallicity\cite{Groot,Parknature}, have sparked tremendous scientific and technological interest. By combining the two fascinating properties, a potentially more interesting state, namely, the Dirac half metal, has been proposed recently based on a model calculation.\cite{Ishizuka} Such a system is not only semiconducting in one spin channel and metallic in the other but also characterized by the zero-gap band structure with a linear dispersion. The coexistence of 100\% spin polarization and massless Dirac fermions makes this system promising for future applications in electronics, spintronics and optoelectronics. In addition, as the Dirac half metal possesses a Dirac cone in only one spin channel (being an ordinary insulator in the other spin-channel), the gap opening triggered by spin-orbit coupling (SOC) leads to topological phase transition in only one spin channel. Hence, the Dirac half metal is also a natural avenue toward the realization of the quantum anomalous Hall effect (QAHE).

Symmetry plays a central role in Dirac materials, such as the sublattice symmetry for graphene and the time-reversal symmetry for topological insulators.\cite{Wehling, CastroRMP, QiToday} It protects two linear dispersive bands crossing at the Dirac point and its breaking leads to a Dirac gap, which is an essential requirement for modern electronics. From this symmetry point of view, the Dirac half metal may be more interesting because there must exist a symmetry order that is destroyed in only one spin channel but not in the other, which thus calls for a distinctly different symmetry property from that of the non-spin-polarized Dirac materials like graphene and topological insulators. Given the great academic interest and potential applications, it is highly desirable to search for experimentally realizable Dirac half metals with simple crystal and electronic structures. As the half metal has broken time-reversal symmetry alongside with a spin-resolved orbital physics, a viable strategy for designing the Dirac half metals is to search for such an orbital physics, whereby the originally degenerate orbitals, e.g., the transition metal (TM) $d$-orbitals, split into two groups with opposite behaviors under the same symmetry operation.

In this work, we introduce the first Dirac half metal system to be engineered through such a mechanism, Mn-intercalated graphene on SiC(0001). By utilizing the substrate modulation, the characteristic Mn $d$-orbitals hybridizing with graphene $p$-orbitals in the two spin channels become different, so that the corresponding $p$-$d$ interaction energies acquire, respectively, odd and even parities under the same quasi-2D inversion operation. Furthermore, as the substrate orders a subset of the Mn atoms, the symmetry-breaking induced gap-opening of the majority spin channel is $\emph{global}$ and the symmetry cannot be recovered through a random distribution or local fluctuations of Mn atoms. Using first-principles calculations, we show that this system becomes a Dirac half metal when the Mn coverage $\chi$ exceeds 1/3 monolayer (ML: Note that 1 ML coverage is defined as one Mn atom per surface Si atom.), having zero gap in the minority spin channel while exhibiting a 150-meV gap in the majority spin channel. The experimental realization of such a system is supported not only by our calculations which indicate that the Dirac half metal is robust against Mn clustering and geometric distortion, but also by previous work which has demonstrated the experimental feasability of Mn intercalation into epitaxial graphene on SiC(0001)\cite{Gao}. Finally, the inclusion of SOC in the calculation reveals that the Dirac half metal exhibits the QAHE, as supported by our Chern number analysis.

Spin-polarized calculations were performed using the Vienna \emph{ab initio} simulation package (VASP) \cite{vasp} within the framework of the density-functional theory (DFT). The local density approximation (LDA)\cite{CA} and the projector-augmented wave \cite{PAW} potential with a cutoff energy of 400 eV were used to describe the exchange-correlation energy and the electron-ion interaction, respectively. A $k$-mesh of 6 $\times$ 6 $\times$ 1 was used to sample the Brillouin zone. The adopted model was to intercalate different coverages of Mn atoms into the epitaxial graphene on SiC(0001), which was based on a $\sqrt3 \times \sqrt3R30^{\circ}$ unit cell for 6$H$-SiC to accommodate a 2 $\times$ 2 graphene. In the simulations, the bottommost three out of the six SiC bilayers were fixed at their respective bulk positions while all other atoms were fully relaxed without any symmetry constraint until the residual forces were less than 0.01 eV/\AA.

\emph{Quasi-2D inversion symmetry.}--The quasi-2D inversion operation is given as follows: $\overrightarrow{\textbf{\emph{OR}}}$ $\rightarrow$ $-\overrightarrow{\textbf{\emph{OR}}}$ and $\overrightarrow{\textbf{\emph{OZ}}}$ $\rightarrow$ $\overrightarrow{\textbf{\emph{OZ}}}$, where \textbf{\emph{O}} is the inversion center of the graphene, $\overrightarrow{\textbf{\emph{OR}}}$  lies in the plane of graphene lattice sites and  $\overrightarrow{\textbf{\emph{OZ}}}$ is normal to the plane (see Fig. 2). A physical quantity possesses the quasi-2D inversion symmetry, only if it is invariant under such an operation. As the graphene Dirac cone is related to its inversion symmetry, introducing the quasi-2D symmetry helps to understand the effect of the TM on the graphene Dirac cone. The Mn-intercalated graphene on SiC(0001) is characterized by the $p$-$d$ hybridization between the $p_z$ oribtals of graphene and the $d$ oribtals of the TM = Mn. Within the Slater-Koster approximation\cite{slater}, the $p$-$d$ interactions exhibit two kinds of symmetry properties, because, upon the quasi-2D inversion, only the in-plane, \emph{\textbf{x}} and \emph{\textbf{y}}-related directional cosines change signs, while the out-of-plane \textbf{\emph{z}}-related directional cosines do not. This leads to a sign change in the $p_z$ (C)-$d_{xz}$ (TM) and $p_z$ (C)-$d_{yz}$(TM) interaction energies, while the signs for the $p_z$ (C)-$d_{xy}$, $p_z$ (C)-$d_{x^2-y^2}$, and $p_z$ (C)-$d_{z^2}$ interaction energies will not change. In other words, the $d_{xz}$ (TM) and $d_{yz}$(TM) orbital interaction with graphene $p_z$ orbitals leads to a perturbation which breaks the inversion symmetry graphene, leading to a Dirac gap.This is not true for the other $d$-orbitals. In this regard, a Dirac half metal could be realized in our system if the $d$-orbitals in each spin channel that participate in the $p$-$d$ hybridization can be adequately engineered.

\begin{figure}
\centerline{\includegraphics[width=.48\textwidth]{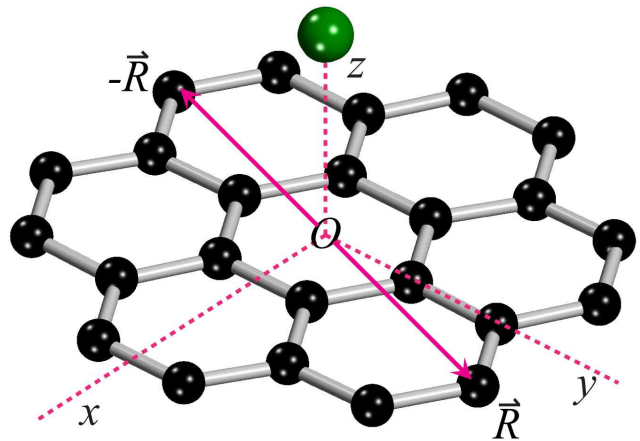}}
\caption{\label{fig:fig1} (Color online)Illustration of the quasi-2D inversion operation to characterize the symmetry physics for a transition metal (Green ball) atom adsorption above the graphene hexagonal center. Under the quasi-2D inversion operation ($\vec{R}\rightarrow-\vec{R}$), the sign changes only for $\textbf{\emph{x}}$ and $\textbf{\emph{y}}$ vectors while not for $\textbf{\emph{z}}$ vector. So do their directional cosines $r_i$ (\emph{i} = $x$, $y$ and $z$). Within the Slater-Koster approximation, the $p$-$d$ interaction energies are $\langle p_z|H|d_{xy}\rangle =r_x r_y r_z(\sqrt{3}V_{pd\sigma}-2V_{pd\pi}$), $\langle p_z|H|d_{x^2-y^2}\rangle =\frac{1}{2}r_z(r^2_x-r^2_y)(\sqrt{3}V_{pd\sigma}-2V_{pd\pi}$), $\langle p_z|H|d_{z^2}\rangle =r_z[(r^2_z-(r^2_x+r^2_y)/2)V_{pd\sigma}+\sqrt{3}(r^2_x+r^2_y)V_{pd\pi}$], $\langle p_z|H|d_{xz}\rangle =r_x[\sqrt{3}r^2_z V_{pd\sigma}+(1-2r^2_z)V_{pd\pi}$], and $\langle p_z|H|d_{yz}\rangle =r_y[\sqrt{3}r^2_z V_{pd\sigma}+(1-2r^2_z)V_{pd\pi}$], respectively, where $V_{pd\sigma}$ and $V_{pd\pi}$ are two parameters. These clearly show that the sign change occurs under the quasi-2D inversion operation for $p_z$ hybridization with either $d_{xz}$ or $d_{yz}$ orbitals, meaning the symmetry breaking. In contrast, they are unchanged for the other three $d$-orbitals ($d_{z^2}$, $d_{xy}$ and $d_{x^2-y^2}$), meaning the symmetry invariant $p$-$d$ interaction.}
\end{figure}

\begin{figure}
\centerline{\includegraphics[width=.48\textwidth]{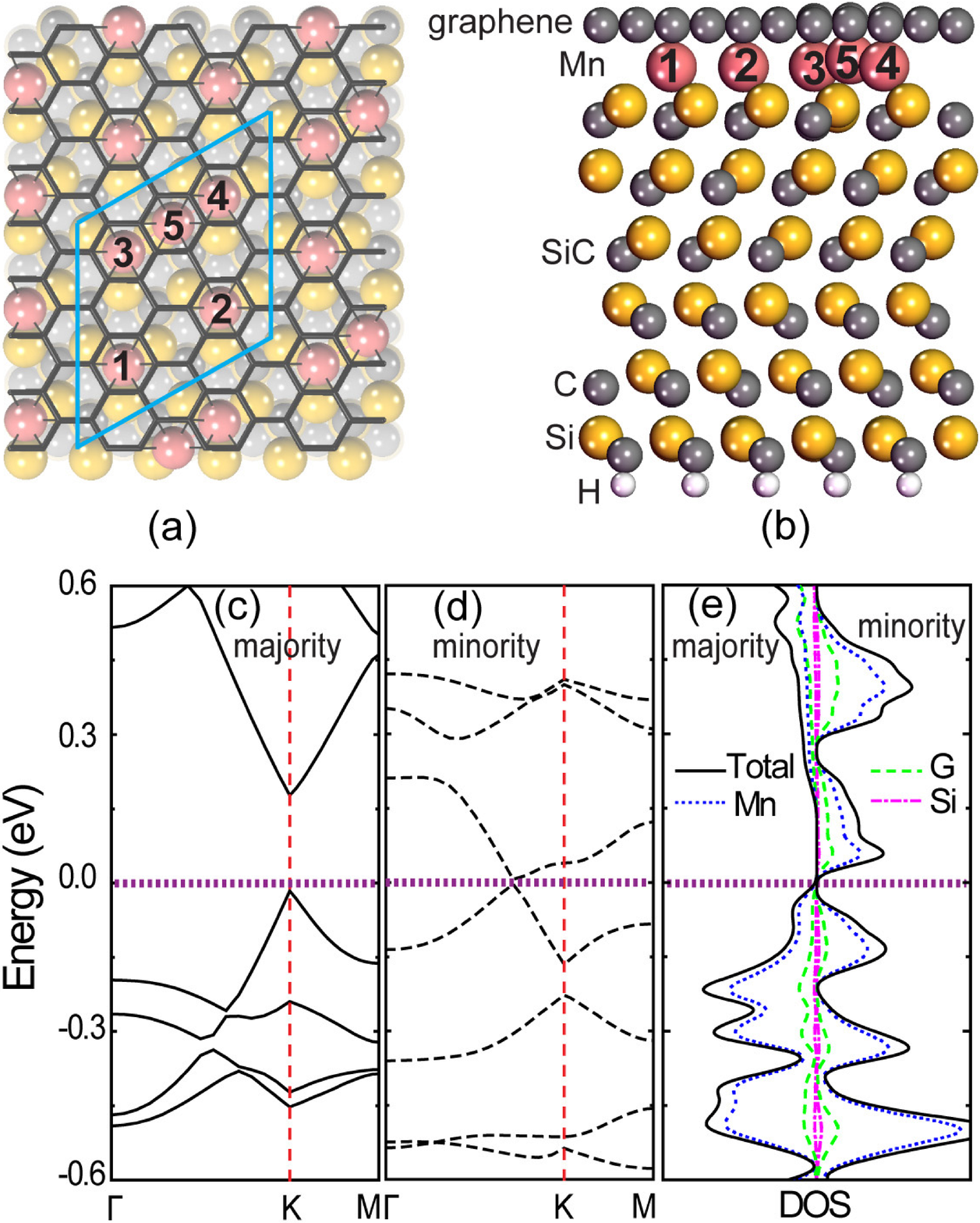}}
\caption{\label{fig:fig2} (Color online)Geometric and electronic structures of Mn intercalated epitaxial graphene on SiC(0001). (a) Top and (b) side views of the optimized geometry for Mn intercalation coverage of $\chi =$ 5/12 ML. Blue rhombus in (a) denotes the supercell. The numbers (1, 2, 3, 4, and 5) mark the five different Mn atoms. Note that there is only one configuration for adding Mn$_5$ into a 2 $\times$ 2 supercell with $\chi$ = 1/3 ML. Here Mn$_3$, Mn$_4$ and Mn$_5$ atoms form a trimer. The corresponding spin-resolved band structure along the high-symmetry lines: (c) majority and (d) minority spin channels. (e) Total density of states as well as the partial contributions from graphene (G), intercalated Mn and surface Si atoms. The Fermi level is set to zero (brown dot). Note that spin-orbit coupling is neglected in this calculation.}
\end{figure}

\emph{Dirac half metal behavior in the Mn intercalated graphene.}--

The $\chi$ = 1/3 ML Mn coverage, corresponding to a single Mn for every three Si at the SiC(0001) surface, is a critical coverage in this system. At $\chi$ = 1/3 ML, each Mn is anchored by three Si and all of the Si dangling bonds (DBs) are passivated and the system is non-magnetic.\cite{usPRL, usPRB} Further insertion of Mn inevitably leads to the formation of Mn trimmers, as illustrated in Figs. 2(a) and 2(b), where the optimized structure for $\chi =$ 5/12 ML is shown. The 5/12 ML coverage is equivalent to adding a fifth Mn atom (Mn$_5$) to a 2 $\times$ 2 supercell with $\chi$ = 1/3 ML.

Figures 2(c) and 2(d) show the spin-resolved band structures for $\chi =$ 5/12 ML. A notable feature is the fully spin-polarized Dirac spectrum. The majority spin channel has an energy gap of 190 meV, while the minority spin channel has a Dirac cone off from $K$. A displacement of the Dirac point to a lower-symmetry point has been frequently seen in the graphene-related systems, e.g., when an anisotropic strain is present.\cite{usPRL,Hasegawa} To confirm this is indeed a Dirac point, we have calculated the three-dimensional band structure (not presented here) to find that the level crossing does happen at the Fermi level. Moreover, by using a 2 $\times$ 1 supercell, we find the ferromagnetic coupling configuration is about 15 meV more stable than the antiferromagnetic one, showing the long range ferromagnetic order.

The Dirac half metal is also revealed by the density of states (DOS) in Fig. 2(e). Here, the majority spin gap is 150 meV, which is smaller than the 190 meV at $K$ in Fig. 2(c), suggesting that the minimum gap may also be off $K$. The projected DOS reveals that the states around the Fermi level are dominated by Mn, indicating its essential role in forming the half-metallicity and linear dispersion. Unlike conventional Dirac systems, here an excited Dirac fermion can be fully spin-polarized with unique advantages in polarization optics and spintronics.\cite{wang}

To explain the first-principles results, let us consider the following two questions: (1) how does the system magnetism arise, and (2) why does the gap open only in the majority-spin channel? Answer to the first question lies in the different interactions of the Mn atoms with the SiC substrate, whereas the answer to the second question lies in the quasi-2D inversion symmetry discussed previously.

For the magnetism, we note that there are only 12 surface Si atoms per unit cell at $\chi =$ 5/12 ML, each with 1 DB. The stable form of Mn on SiC(0001) is to bind with 3 Si DBs in a $T_4$ configuration, as we have demonstrated before\cite{usPRL}. This means at 1/3 ML, all the DBs have been saturated. The lack of sufficient Si DBs at 5/12 ML naturally divides the Mn atoms into two categories: (i) Mn$_1$-Mn$_4$, each is chemically bound to three Si atoms and (ii) Mn$_5$, which is atom-like because of no more surface Si to bind to, in spite of that the crystal field effect promotes all its 7 valence electrons into the $3d$ orbitals. For category-(i) Mn, binding to Si completely quenches its spin-polarization\cite{usPRL}. Among the crystal-field-split $d$-orbitals, the $z$-involving orbitals, $d_{xz}$, $d_{yz}$, and $d_{z^2}$, are strongly hybridized with the Si DBs to result in significantly lower energies, as shown in Fig. 3(a). The remaining $d_{xy}$ and $d_{x^2-y^2}$ orbitals are, however, little affected by the Si from the substrate and hence they maintain their non-bonding characters with energies near the Dirac point of the graphene. For category-(ii) Mn, on the other hand, as none of its $d$-orbitals can be significantly affected by the substrate, they are all atom-like, i.e., non-bonding with energies near the Dirac point of the graphene, as shown in Fig. 3(b). In this case, the 7 $d$-electrons should occupy 5 spin-up states and 2 spin-down states to result in a high spin configuration of 3 $\mu_B$, as confirmed by our first-principles calculation.

\begin{figure}
\centerline{\includegraphics[width=.48\textwidth]{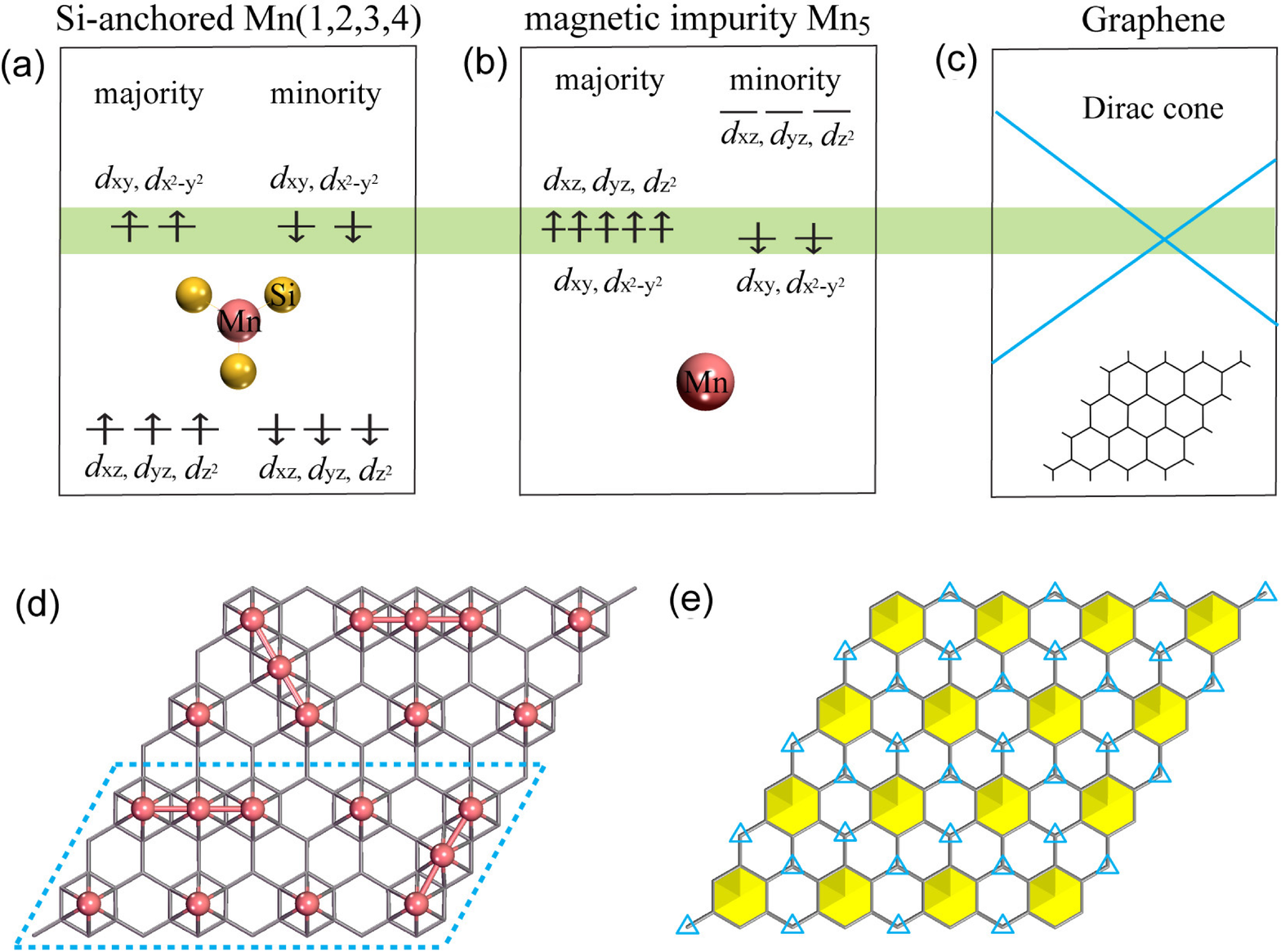}}
\caption{\label{fig:fig3}  (Color online)The $d$-electron occupations of two categories of Mn atoms and the schematic diagram of their electronic levels with respect to graphene Dirac point. (a) Mn$_1$ to Mn$_4$, fully saturated by three Si DBs. (b) Mn$_5$, atomic-like magnetic impurity. (c) Graphene Dirac cone. The relative energy alignments are deduced from the symmetry analysis in combination with our DFT calculations. Green box denotes those states important for the $p$-$d$ coupling. (d) Schematic illustration of the disorder due to possibly random distribution of Mn$_5$. For clarity, only graphene and Mn (pink balls) are shown. Blue rhomboid denotes the 2 $\times$ 1 supercell containing two Mn trimers used for test calculations on random disorder. (e) Geometry at $\chi =$ 1/3 ML, where there exist category (i) Mn atoms underneath the yellow hexagons. Because of this pre-established distribution of Mn, the symmetry breaking cannot be canceled on average and only the one on the sites (in triangle) bonding-free to category (i) Mn is anticipated to be suppressed by the random effect.}
\end{figure}

For the gap opening, we consider the coupling between Mn $d$ and graphene $p_z$ orbitals. For substantial coupling to exists, the states need to be close in energy, namely, between the non-bonding $d$ orbitals highlighted in the green areas in Figs. 3(a) and 3(b) and $p_z$ orbitals in Fig. 3(c). For the spin-minority channel, the coupled states only involve $d_{xy}$ and $d_{x^2-y^2}$, whose hybridization with $p_z$ orbitals holds the quasi-2D inversion symmetry, which protects the Dirac cone. In contrast in the spin-majority channel, the symmetry-breaking interactions involving the $d_{xz}$ and $d_{yz}$ orbitals open a Dirac gap. Put together, this analysis shows that the Dirac half metal arises due to the occupation-modulation effect by SiC substrate on the Mn $d$ orbitals and the quasi-2D inversion symmetry of the combined graphene-plus-Mn system, as shown in Fig. 1.

We calculate the Mn intercalation energy, which is defined as the total energy difference between (Mn bulk + epitaxial graphene on SiC) and the Mn intercalated system. We obtain 0.4, 0.3, and 0.0 eV/Mn, for $\chi =$ 1/3, 5/12, and 1/2 ML, respectively. The higher stability at low $\chi$ suggests that Mn clustering is energetically unfavorable. In other words, Mn tends to avoid each other, which allows them to maximize the binding to available Si sites on the substrate. Thus, trimer formation primarily happens after all Si sites have been fully passivated, i.e., 1/3 ML.

This means only when $\chi >$ 1/3 ML, there emerges the category-(ii) Mn and the system becomes a Dirac half metal. As $\chi$ gradually increases to 5/12 ML, more and more Mn trimers form, and the system remains a Dirac half metal according to our analysis based on the two categories of Mn. This is explicitly corroborated by our first-principles calculations on the $\chi=$ 17/48 ML system (containing only one Mn trimer in a $4\sqrt3 \times 4\sqrt3R30^{\circ}$ SiC lattice). Increasing the Mn coverage so that all of the Mn atoms form trimers ($\chi =$ 1/2 ML), however, results in a system which is no longer a Dirac half metal, and instead exhibits charge transfer from the majority to minority spin channel and a non-integer total moment of 2.8 $\mu_B$. These calculations indicate that the Mn intercalated SiC (0001) system possesses a wide range of coverages (1/3 ML$< \chi<\chi_{\max}$), with (5/12 ML $< \chi_{\max} <$ 1/2 ML),  in which it exists in the Dirac half metal state.

The use of supercells in these calculations raises the important question as to whether the effects of the symmetry breaking will diminish, or be completely eliminated, in realistic systems in which the Mn trimers may distribute randomly, as schematically shown in Fig. 3(d). This is indeed the case for systems in which the symmetry is only locally broken, and the averaging of the perturbations leads to a situation which preserves the bulk symmetry of the graphene on substrate\cite{Bostwick,Hass,Varchon,Ortix}. While our first-principles calculations suggest that the gap persists in the presence of disorder (e.g. the 2 $\times$ 1 supercell [blue rhomboid in Fig. 3(d)] with reduced order is not only $\sim$13 meV less stable than the one in Fig. 2(a) but still yields a similar half metallic gap), this is not a question which can be answered, strictly speaking, based on a supercell approximation which imposes long range order. The important realization is that the substrate fixes the positions of category-(i) Mn [Mn$_1$-Mn$_4$] and as a result the symmetry breaking perturbation of category-(ii) Mn [Mn$_5$] \emph{cannot} be completely random.

To illustrate this last point, consider the worst case scenario for maintaining a symmetry breaking perturbation, namely, complete randomness of the distribution of Mn$_5$ atoms on all of the possible Mn$_5$ sites (centered on the white hexagons in Fig. 3(e)). If one adds all of the asymmetric Slater-Koster interaction terms effecting graphene resulting from such a distribution, one finds that they completely cancel only on the C-atoms marked with a triangle in Fig. 3(e). Note that all of the C-atoms adjacent to a category-(i) Mn [Mn$_1$-Mn$_4$, centered on the yellow hexagons] are still effected by symmetry breaking perturbations. The reason complete cancellation does not occur is because Mn$_5$ cannot occupy the yellow hexagon sites. This regular lattice of sites is determined by the substrate and any Mn which sits at one of these sites is of category-(i) due to its interaction with the substrate. As a result, even a completely ``random" distribution of Mn$_5$ still has long range order in that there is a regular lattice of sites from which it is excluded. Hence, the symmetry breaking of this system is inherently global and cannot be eliminated due to randomness in the distribution of Mn$_5$.

\begin{figure}
\centerline{\includegraphics[width=.48\textwidth]{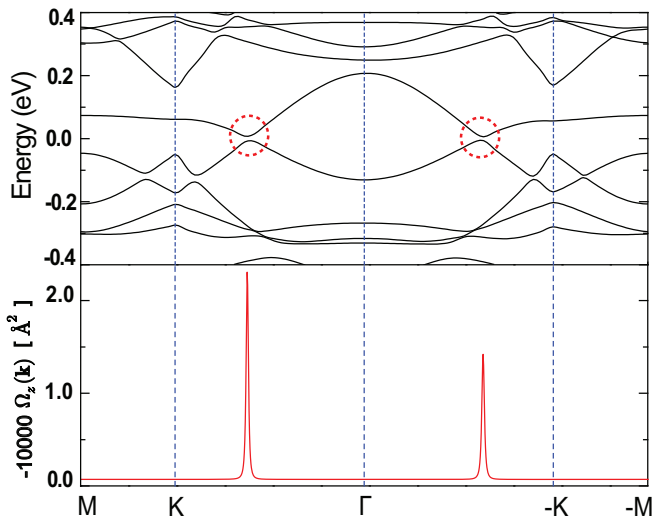}}
\caption{\label{fig:fig4} (Color online) Band structure including the SOC (top panel) and the Berry curvature (bottom panel) along the high-symmetry line of Mn intercalated epitaxial graphene on SiC(0001) at $\chi =$ 5/12 ML. Red circles highlight the SOC gap. The Fermi level is set to zero.}
\end{figure}

\emph{Quantum anomalous Hall effect.}--It has been shown previously that Mn intercalation significantly enhances the SOC. At $\chi=$ 1/3 ML, it opens a band gap of 26 meV. This drives the system into a quantum spin Hall state.\cite{usPRB} In the present case, Mn$_5$ makes the system spin-polarized, thereby breaking the time reversal symmetry. In this case, the SOC should open the Dirac cone in the minority spin channel with a nontrivial gap and thus the system enters a QAH state. First-principles calculations including the SOC indeed yields a SOC gap around 20 meV at $\chi=$ 5/12 ML, as shown in the top panel of Fig. 4 (Red circles). To identify the topological properties of the gapped state, we employ the Chern number analysis\cite{chern,WangX} in which the Chern number is calculated by a \emph{k}-space integral over the first Brillouin zone, $C = \frac{1}{2\pi}\int_{\rm BZ}\Omega(\rm \textbf{k})d^2\textbf{k}$, where $\Omega(\rm \textbf{k})$ is the Berry curvature of all occupied states\cite{Berry}:
\begin{equation}\label{(1)}
\Omega(\textbf{k})=-\sum_{n<E_F}\sum_{m\neq n} {\rm 2Im}\frac{\langle\psi_{n\textbf{k}}|\nu_x|\psi_{m\textbf{k}}\rangle\langle\psi_{m\textbf{k}}|\nu_y|\psi_{n\textbf{k}}\rangle}{(\varepsilon_{m\textbf{k}}-\varepsilon_{n\textbf{k}})^2}.
\end{equation}
$\psi_{n \rm \textbf{k}}$ is the spinor Bloch wave function of band $n$ with corresponding eigenenergy $\varepsilon_{n \rm \textbf{k}}$, and $\nu_i$ is the \emph{i}th Cartesian component of the velocity operator. The calculations are conducted by using the WANNIER90 package\cite{WangX,wannier} with the maximally localized Wannier functions, which reproduces the DFT results well (not shown). Figure 4 (bottom panel) shows the calculated total Berry curvature along the high symmetry lines. The large peaks between $\Gamma$ and $K$ ($-K$) points arise where the intersections between the conduction and valance bands split by the SOC, giving rise to small denominators in Eq. (1), and hence a large contribution to $\Omega(\rm \textbf{k})$. By integration over the Brillouin zone, we obtain an odd Chern number, $C=1$, to confirm that, in a ribbon structure, only one chiral state can exist on the edge. It should be noted that this Chern number is different from that in TM-doped graphene\cite{ZhangHB,QiaoZH} and silicene\cite{ZhangXL} where the Chern number is even ($C=2$).

To summarize, we propose a strategy to make Dirac half metals in experimentally-accessible Mn intercalated epitaxial graphene on SiC(0001), whose low-energy excitations have a spin-gapless Dirac spectrum, originated from the coupling between the substrate-modulated Mn $d$ orbitals and graphene $p_z$ orbitals. In the majority spin channel, the interactions involving Mn $d_{xz}$ and $d_{yz}$ orbitals break the quasi-2D inversion symmetry and thus induce a Dirac gap. In the minority spin channel, however, there exist only $d_{xy}$ and $d_{x^2-y^2}$ interactions, which leaves the Dirac cone intact. Furthermore, the substantial SOC of Mn drives the Dirac half metal system into a QAH state. As the SOC is an intrinsic property of the material, the QAHE is also an intrinsic property of the Dirac half metal. We thus anticipate that our study opens a new route toward the realization of the QAHE.

\begin{acknowledgments} We thank Hongbin Zhang and David Vanderbilt for valuable discussions. This work was supported by the Ministry of Science and Technology of China (Grant Nos. 2011CB606405, 2011CB921901 and 2009CB929400), the National Natural Science Foundation of China (Grant Nos. 11304053, 11104155 and 11334006), and Open Research Fund Program of the State Key Laboratory of Low-Dimensional Quantum Physics. The work at RPI of D. West was supported by the Defense Advanced Research Project Agency (DARPA), Award No. N66001-12-1-4034, and that of S.B. Zhang by the Department of Energy under Grant No. DE-SC0002623.
\end{acknowledgments}

%\end{bibliography}

\end{document}